\documentclass[aps,prc,twocolumn,superscriptaddress,nofootinbib,showpacs,showkeys]{revtex4-1}
\usepackage{graphicx}  
\usepackage{dcolumn}   
\usepackage{bm}        
\usepackage{amssymb}   
\usepackage{amsmath}   
\usepackage{lineno}
\usepackage{draftcopy}
\usepackage{relsize} 
\usepackage{xfrac}    

\usepackage{color}
\definecolor{dgreen}{cmyk}{1.,0.,1.,0.2}        
\definecolor{orange}{cmyk}{0.,0.353,1.,0.}    

\usepackage[bookmarks]{hyperref}

\newcommand{ \bp }{{\bf p}}
\newcommand{ \bv }{{\bf v}}

\newcommand{ \bmu }{{\boldsymbol \mu}}

\newcommand{ \bB }{{\bf B}}

\newcommand{ \bP }{{\bf P}}

\newcommand{ \bS }{{\bf S}}

\newcommand{ \bomega }{{\boldsymbol \omega}}
\newcommand{ \bOmega }{{\boldsymbol \Omega}}
\newcommand{\tomega}{\varpi}

\newcommand{ \Lambdabar }{{\bar{\Lambda}}}

\newcommand{ \mean }[1]{\left\langle #1 \right\rangle}

\newcommand{ \psioep }{\Psi_{\rm EP}^{(1)}}

\newcommand{\bra}[1]{\langle #1|}
\newcommand{\ket}[1]{|#1\rangle}
\newcommand{\braket}[2]{\langle #1|#2\rangle}

\newcommand{\di}{{\rm d}}
\newcommand{\ii}{i}

\def\wT{{\widehat T}}

\def\wQ{{\widehat Q}}

\def\wrho{{\widehat{\rho}}}

\newcommand{\tr}{{\rm tr}}

\newcommand{\omegav}{\boldsymbol{\omega}}

\newcommand{\p}{{\rm p}}

\newcommand{\be}{\begin{equation}}
\newcommand{\ee}{\end{equation}}                                                                               
\newcommand{\bea}{\begin{eqnarray}}
\newcommand{\eea}{\end{eqnarray}}

\begin{document}

\setlength{\linenumbersep}{6pt}

\title{
Global hyperon polarization at local thermodynamic equilibrium with
vorticity, magnetic field and feed-down}
\author{Francesco Becattini} \affiliation{Dipartimento di Fisica,
  Universita` di Firenze, and INFN, Sezione di Firenze, Florence,
  Italy} 
\author{Iurii Karpenko} \affiliation{INFN, Sezione di Firenze, Florence,
  Italy} 
\author{Michael Annan Lisa} \affiliation{Physics Department, The Ohio
  State University, Columbus, Ohio 43210, USA} 
\author{Isaac Upsal} \affiliation{Physics Department, The Ohio State
  University, Columbus, Ohio 43210, USA}
\author{Sergei A. Voloshin} \affiliation{Wayne State University,
  Detroit, Michigan 48201, USA}

\begin{abstract}
The system created in ultrarelativistic nuclear collisions is known to behave as an almost ideal liquid. 
In non-central collisions, due to the large orbital momentum, such a system might be the fluid with 
the highest vorticity ever created under laboratory conditions. Particles emerging from such a 
highly vorticous fluid are expected to be globally polarized with their spins on average pointing 
along the system angular momentum. Vorticity-induced polarization is the same for particles and 
antiparticles, but the intense magnetic field generated in these collisions may lead to the 
splitting in polarization. In this paper we outline the thermal approach to the calculation of
the global polarization phenomenon for particles with spin and we discuss the details of the 
experimental study of this phenomenon, estimating the effect of feed-down. A general formula 
is derived for the polarization transfer in two-body decays and, particularly, for strong and
electromagnetic decays. We find that accounting for such effects is crucial when extracting vorticity 
and magnetic field from the experimental data.

\end{abstract}


\maketitle

\section{Introduction}

Heavy ion collisions at ultrarelativistic energies create a strongly
interacting system characterized by extremely high temperature and
energy density. For a large fraction of its lifetime the system shows
strong collective effects and can be described by relativistic
hydrodynamics. In particular, the large elliptic flow observed in
such collisions, indicate that the system is strongly coupled, with
extremely low viscosity to entropy ratio~\cite{Heinz:2013th}. From the
very success of the hydrodynamic description, one can also conclude
that the system might possess an extremely high vorticity, likely the
highest ever made under the laboratory conditions.

\begin{figure}
\includegraphics[keepaspectratio, width=1.\columnwidth]{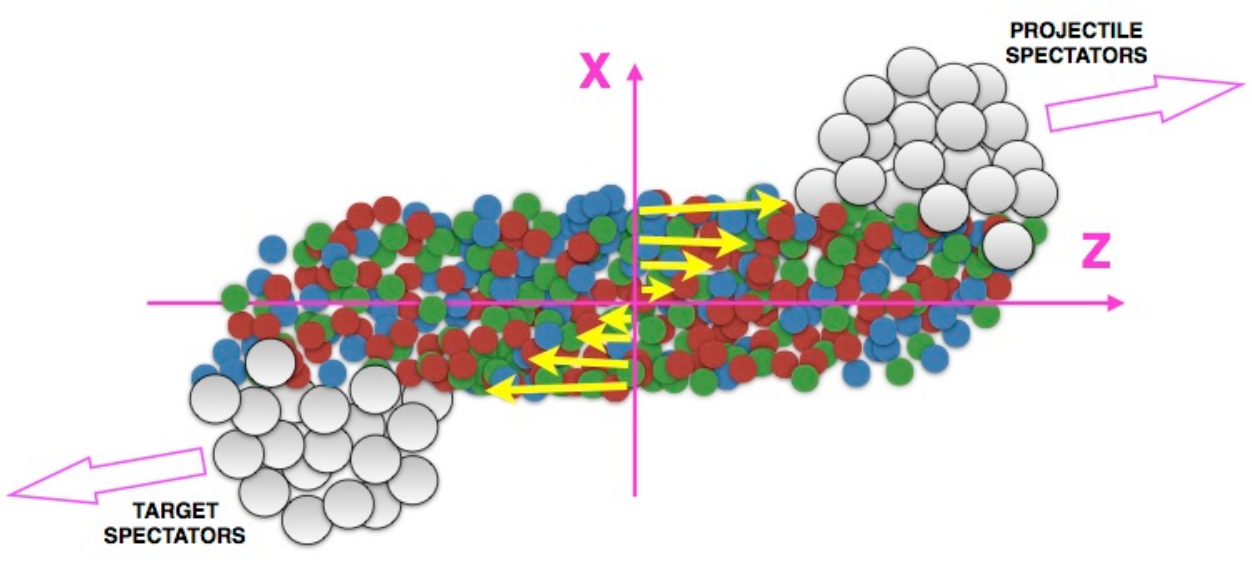}
  \caption{
    Schematic view of the collision. Arrows indicate the flow velocity
  field.
  The $+\hat{y}$ direction is out of the page; both
  the orbital angular momentum and the magnetic field point
  into the page.}
  \label{fig:collision}
\end{figure}

A simple estimate of the non-relativistic vorticity, defined
as\footnote{sometimes the vorticity is defined without the factor
  $1/2$; we use the definition that gives the vorticity of the fluid
  rotating as a whole with a constant angular velocity $\Omega$, to be
  $\omega=\Omega$}
\be\label{nrv}
 \bomega =\frac{1}{2}\, \nabla \times \bv,
\ee
can be made based on a very schematic picture of the collision
depicted in Fig.~\ref{fig:collision}. As the projectile and target
spectators move in opposite direction with the velocity close to the
speed of light, the $z$ component of the collective velocity in the
system close to the projectile spectators and that close to the target
spectators are expected to be different. Assuming that this difference
is a fraction of the speed of light, e.g.  0.1 (in units of the speed
of light), and that the transverse size of the system is about 5~fm,
one concludes that the vorticity in the system is of the order
$0.02\,{\rm fm^{-1}}\approx 10^{22}\,{\rm s}^{-1}$.  

In relativistic hydrodynamics, several extensions of the
non-relativistic vorticity defined above can be introduced (see
ref.~\cite{Becattini:2015ska} for a review).  As we will see below,
the appropriate relativistic quantity for the study of global
polarization is the {\it thermal vorticity}:
\begin{equation}\label{thvort}
  \tomega_{\mu\nu} = 
\frac{1}{2} \left( \partial_\nu \beta_\mu - \partial_\mu \beta_\nu \right)
\end{equation}
where $\beta_\mu = (1/T) u_\mu$ is the four-temperature vector, $u$
being the hydrodynamic velocity and $T$ the proper temperature. At an
approximately constant temperature, thermal vorticity can be roughly
estimated by $\tomega \sim \omega/T~$ where $\omega$ is the local
vorticity, which, for typical conditions, appears to be of the order
of a percent by using the above estimated vorticity and the
temperature $T \sim 100$~MeV.

Vorticity plays a very important role in the system evolution. Accounting for 
vorticity (via tuning the initial conditions and specific viscosity) it was possible 
to quantitatively describe the rapidity dependence of directed
flow~\cite{Abelev:2013cva,Adare:2015cpn}, which, at present, can not
be described by any model not including initial angular
momentum~\cite{Csernai:2013bqa,Bozek:2010bi,Becattini:2015ska}.

Vorticous effects may also strongly affect the baryon dynamics of the
system, leading to a separation of baryon and antibaryons along the
vorticity direction (perpendicular to the reaction plane) -- the
so-called Chiral Vortical Effect (CVE). The CVE is similar in many
aspects to the more familiar Chiral Magnetic Effect (CME) - the
electric charge separation along the magnetic field. For recent
reviews on those and similar effects, as well as the status of the
experimental search for those phenomena,
see~\cite{Kharzeev:2015znc,Skokov:2016yrj}. For a reliable theoretical
calculation of both effects one has to know the vorticity of the
created system as well as the evolution of (electro)magnetic field.

Finally, and most relevant for the present work, vorticity induces a
local alignment of particles spin along its direction. The general
idea that particles are polarized in peripheral relativistic heavy ion
collisions along the initial (large) angular momentum of the plasma
and its qualitative features were put forward more than a decade
ago~\cite{Liang:2004ph,Voloshin:2004ha,
  Abelev:2007zk,Gao:2007bc,Betz:2007kg}. The idea that polarization is
determined by the condition of local thermodynamic equilibrium and its
quantitative link to thermal vorticity were developed in
refs.~\cite{Becattini:2007nd,Becattini:2013vja}. The assumption that
spin degrees of freedom locally equilibrate in much the same way as
momentum degrees of freedom makes it possible to provide a definite
quantitative estimate of polarization through a suitable extension of
the well known Cooper-Frye formula.

This phenomenon of global (that is, along the common direction of the
total angular momentum) polarization has an intimate relation to the
Barnett effect~\cite{Barnett:1915} - magnetization by rotation - where
a fraction of the orbital momentum associated with the body rotation
is irreversibly transformed into the spin angular momentum of the
atoms (electrons), which, on the average, point along the angular
vector. Because of the proportionality between spin and magnetic
moment, this tiny polarization gives rise to a finite magnetization of
the rotating body, hence a magnetic field. Even closer to our case is
the recent observation of the electron spin polarization in vorticous
fluid~\cite{Takahashi:2016} where the ''global polarization'' of
electron spin has been observed due to non-zero vorticity of the
fluid. In condensed matter physics the gyromagnetic phenomena are
often discussed on the basis of the so-called Larmor's
theorem~\cite{Heims:1962}, which states that the effect of the
rotation on the system is equivalent to the application of the
magnetic field $\bB=-\gamma^{-1} \bOmega$, where $\gamma$ is the
particle gyromagnetic ratio.

It is worth pointing out that, however, polarization by rotation and
by application of an external magnetic field are conceptually distinct
effects. Particularly, the polarization by rotation is the same for
particles and antiparticles, whereas polarization by magnetic field is
opposite. This means that, for example, magnetization by rotation
(i.e. Barnett effect) cannot be observed in a completely neutral
system and the aforementioned Larmor's theorem cannot be applied; for
this purpose, an imbalance between matter and antimatter is necessary.

In this regard, the global polarization phenomenon in heavy ion
collisions is peculiarly different from that observed in condensed
matter physics for the density of particles and antiparticles are
approximately equal, so that non-zero global polarization does not
necessarily imply a magnetization. This system thus provides a unique
possibility for a direct observation of the transformation of the
orbital momentum into spin. Furthermore, note that in heavy ion
collisions, the polarization of the particles can be directly measured
via their decays (in particular via parity violating weak decays).

Calculations of global polarization in relativistic heavy ion
collisions have been performed using different techniques and
assumptions. Several recent calculations employ 3+1D hydrodynamic
simulations and use the assumption of local thermodynamic
equilibrium~\cite{Becattini:2013vja,Becattini:2015ska,
  Pang:2016igs,becakarp}; observing quite a strong dependence on the
initial conditions. While local thermodynamic equilibrium for the spin
degrees of freedom remains an assumption - as no estimates of the
corresponding relaxation times exist - such an approach has a clear
advantage in terms of simplicity of the calculations. All of the
discussion below is mostly based on this assumption; to simplify the
discussion even more, we will often use the non-relativistic limit.

It should be pointed out that different approaches - without local
thermodynamic equilibrium - to the estimate of $\Lambda$ polarization
in relativistic nuclear collisions were also proposed~\cite{ayala,
  celso,Baznat:2015eca,Aristova:2016wxe}.
 
The paper is organized as follows: in Section~\ref{poladef} we
introduce the main definitions concerning spin and polarization in a
relativistic framework; in Section~\ref{thermal} we outline the
thermodynamic approach to the calculation of the polarization and
provide the relevant formulae for relativistic nuclear collisions; in
Section~\ref{lambda} we address the measurement of $\Lambda$
polarization and in Section~\ref{align} the alignment of vector
mesons; finally in Section~\ref{feeddown} we discuss in detail the
effect of decays on the measurement of $\Lambda$ polarization.

\subsection*{Notation}

In this paper we use the natural units, with $\hbar=c=k_B=1$.
The Minkowskian metric tensor is ${\rm diag}(1,-1,-1,-1)$; for the Levi-Civita
symbol we use the convention $\epsilon^{0123}=1$. Operators in 
Hilbert space will be denoted by a large upper hat, e.g. $\wT$ while unit 
vectors with a small upper hat, e.g. $\hat v$.
 
\section{Spin and polarization: basic definitions}\label{poladef}

In non-relativistic quantum-mechanics, the mean spin vector is defined
as:
\be\label{spin1}
  {\bf S} = \langle \widehat{\bf S} \rangle = \tr (\wrho \, \widehat{\bf S})
\ee
where $\wrho$ is the density operator of the particle under
consideration and $\widehat{\bf S}$ the spin operator. The density
operator can be either a pure quantum state or a mixed state, like in
the case of thermodynamic equilibrium. The polarization vector is
defined as the mean value of the spin operator normalized to the spin
of the particle:
\be\label{spin2}
  {\bf P} = \langle \widehat{\bf S} \rangle/S 
\ee
so that its maximal value is 1, that is $\| {\bf P} \| \le 1$.  

A proper relativistic extension of the spin concept, for massive
particles, requires the introduction of a spin four-vector
operator. This is defined as follows (see e.g. \cite{wukitung}):
\begin{equation}
 \widehat{S}^\mu = -\frac{1}{2m} 
\epsilon^{\mu\nu\rho\lambda} \widehat{J}_{\nu\rho} \widehat{p}_\lambda
\end{equation}
where $\widehat{J}$ and $\widehat{p}$ are the angular momentum
operator and four-momentum operator of a single particle. As it can
be easily shown, the spin four-vector operator commutes with the
four-momentum operator (hence it is a compatible observable) and it is
space-like on free particle states as it is orthogonal to the
four-momentum:
\be
  \widehat{S}^\mu \widehat{p}_\mu = 0
\ee
and has thus only three independent components. Particularly, in the
rest frame of the particle, it has vanishing time component. Because
of these properties, for single particle states with definite
four-momentum $p$ it can be decomposed \cite{moussa} along three
spacelike vectors $n_i(p)$ with $i=1,2,3$ orthogonal to $p$:
\begin{equation}\label{spindec}
  \widehat{S}^\mu = \sum_{i=1}^3 \widehat{S}_i(p) n_i(p)^\mu
\end{equation}
It can be shown that the operators $\widehat{S}_i(p)$ with $i=1,2,3$
obey the well known SU(2) commutation relations and they are indeed
the generators of the little group, the group of transformations
leaving $p$ invariant for a massive particle. Furthermore, it is worth
pointing out that $\widehat{S}^\mu\widehat{S}_\mu$ operator commutes
with both momentum and spin (it is a Casimir of the full Poincar\'e
group) and takes on the value $S(S+1)$ where $S$ is {\em the} spin of
the particle over all states.

The spin and polarization four-vectors can now be defined by a
straighforward extension of the eqs.~(\ref{spin1}), (\ref{spin2}),
namely:
\be\label{spinv}
 S^\mu = \langle \widehat{S}^\mu \rangle \equiv \tr (\wrho \, \widehat{S}^\mu )
\ee
and
\be\label{polav}
 P^\mu = \langle \widehat{S}^\mu \rangle/ S
\ee
In the particle rest frame, both four-vectors have vanishing time
component and effectively reduce to three-vectors. Henceforth, they
will be denoted with an asterisk, that is:
\be
  S^* = (0, {\bf S}^*) \qquad P^* = (0, {\bf P}^*)
\ee
Obviously, they will have non-trivial transformation relations among
different inertial frames, unlike in non-relativistic quantum
mechanics where they are simply invariant under a galilean
transformation.

For an assembly of particles, or in relativistic quantum field theory, the mean 
single-particle spin vector of a particle with momentum $p$ can be written:
\be\label{meanp}
  S^\mu(p) = -\frac{1}{2m} \epsilon^{\mu\nu\rho\lambda} \frac{\sum_{\sigma} 
 \tr \left( \wrho \, \widehat{J}_{\nu\rho} \widehat{p}_\lambda a^\dagger_{p,\sigma}
 a_{p,\sigma} \right)}{ \sum_\sigma \tr (\wrho \, a^\dagger_{p,\sigma} a_{p,\sigma}) }
\ee
where $\wrho$ is the density operator, $\widehat J$ and $\widehat p$
are the {\em total} angular momentum and four-momentum operators,
$a_{\sigma,p}$ is the destruction operator of a particle with momentum
$p$ and spin component (or helicity) $\sigma$.

\section{The thermal approach}\label{thermal}

\subsection{Non-relativistic limit}

Suppose we have a non-relativistic particle at equilibrium in a
thermal bath at temperature $T$ in a rotating vessel at an angular
velocity $\omegav$ (corresponding to a uniform vorticity according to
eq.~(\ref{nrv}) and we want to calculate its mean spin vector
according to eq.~(\ref{spin1}). As spin is quantum, we have to use the
appropriate density operator $\wrho$ for this system at equilibrium,
that in this case reads \cite{Landau:v5,Vilenkin:1980zv}:
\bea\label{thermrho}
  \wrho &=& \frac{1}{Z} \exp[-\widehat{H}/T + \nu \wQ/T + \omegav \cdot \widehat{\bf J}/T 
   + \widehat\bmu \cdot \bB/T] \nonumber \\
  &=& \frac{1}{Z} \exp[-\widehat{H}/T + \nu \wQ/T +\omegav \cdot (\widehat{\bf L} + 
  \widehat{\bf S})/T + \widehat\bmu \cdot \bB/T]]  \nonumber \\
\eea
where for completeness we have included a conserved charges $\wQ$
($\nu$ being the corresponding chemical potentials) and a constant and
uniform external magnetic field $\bB$ ($\widehat \bmu = \mu
\widehat\bS/S$ being the magnetic moment). Indeed, the angular
velocity $\omegav$ plays the role of a chemical potential for the
angular momentum and particularly for the spin. If the constant angular velocity
$\omegav$, as well as the constant magnetic field $\bB$ are parallel, the above
density operator can be diagonalized in the basis of eigenvectors of
the spin operator component parallel to $\omegav$,
$\widehat{\bS}\cdot\hat\omegav$, thereby giving rise to a probability
distribution for its different eigenvalues $m$. Specifically, the different 
probabilities read:
\bea\label{proba}
 w[T,B,\omega](m) = \frac{\exp \left[ \frac{\mu B/S+\omega}{T} m \right]}
 {\sum_{m=-S}^S \exp \left[ \frac{\mu B/S+\omega}{T} m \right]}
\eea
The distribution eq.~(\ref{proba}) may now be used to estimate the spin
vector in eq.~(\ref{spin1}). Indeed, the only non-vanishing component
of the spin vector is along the angular velocity direction; for the
simpler case with $B=0$ this reads:
\bea\label{spin3} 
\bS &=& \hat\omegav \frac{\sum_{m=-S}^S m \exp \left[ \frac{\omega}{T} m \right]}
 {\sum_{m=-S}^S \exp \left[ \frac{\omega}{T} m \right]} \nonumber \\
  &=& \hat\omegav \frac{\partial}{\partial (\omega/T)} 
  \sum_{m=-S}^S \exp \left[ \frac{\omega}{T} m \right] \nonumber \\
 &=& \hat\omegav \frac{\partial}{\partial (\omega/T)} \frac{\sinh[(S+1/2)\omega/T]}{\sinh[\omega/2T]}
\eea
where $\hat\omegav$ is the unit vector along the direction of
$\omegav$. In most circumstances, (relativistic heavy ion collisions
as well) the ratio between $\omega$ and $T$ is very small and a first
order expansion of the above expressions turns out to be a very good
approximation.  Thus, the eq.~(\ref{spin3}) becomes:
\be\label{spin4}
  \bS \simeq \hat\omegav \frac{\sum_{m=-S}^S m^2 \omega/T}{2S+1} = 
  \frac{S(S+1)}{3}\frac{\omegav}{T}
\ee
We can now specify the polarization vector for the particles with
lowest spins. For $S=1/2$ the eqs.~(\ref{spin3}) and (\ref{spin4})
imply:
\be
\label{eq:SpinOneHalf}
 \bS = \frac{1}{2} {\bf P} = \frac{1}{2} \tanh(\omega/2T) \hat\omegav \simeq \frac{1}{4} \frac{\omegav}{T} \; ; 
\ee
for $S=1$:
\be\label{eq:s2}
 \bS = {\bf P} = \frac{2\sinh(\omega/T)}{1+2\cosh(\omega/T)} \hat\omegav 
 \simeq \frac{2}{3} \frac{\omegav}{T}  \; ;  
\ee 
and finally, for $S=3/2$:
\bea\label{eq:s3}
&&  \bS = \frac{3}{2} {\bf P} \nonumber \\
&& = \frac{(3/2)\sinh(3\omega/2T)+(1/2)\sinh(\omega/2T)}{\cosh(3\omega/2T)+\cosh(\omega/2T)} 
 \hat\omegav \simeq \frac{5}{4}\frac{\omegav}{T} \; .
\eea
If the magnetic field is parallel to the vorticity, magnetic effects may be included by substituting:
\begin{equation}
\omegav \rightarrow\omegav +\mu {\bf B}/S
\end{equation}
in equations~(\ref{spin3}-\ref{eq:s3}).

\subsection{Relativistic case}

As it has been mentioned at the beginning of this section, all above
formulae apply to the case of an individual (i.e. Boltzmann
statistics) non-relativistic particle at global thermodynamic
equilibrium with a constant temperature, uniform angular velocity and
magnetic field. It therefore must be a good approximation when the
physical conditions are not far from those, namely a non- relativistic
fluid made of non-relativistic particles with a slowly varying
temperature, vorticity (\ref{nrv}) and magnetic field. However, at
least in relativistic nuclear collisions, the fluid velocity is
relativistic, massive particles with spin may be produced with momenta
comparable to their mass, and the local relativistic vorticity -
whatever it is - may not be uniform.  Furthermore, there is a general
issue of what is the proper relativistic extension of the angular
velocity or the ratio $\omega/T$ appearing in all above formulae. The
fully relativistic ideal gas with spin, in the Boltzmann
approximation, at global equilibrium with rotation was studied in
detail in refs.~\cite{Becattini:2007nd,Becattini:2009wh}. Therein, it
was found that the spin vector in the rest frame, for a particle with
spin $S$ is given by:
\bea\label{genspin}
 && \bS^* =S{\bf P}^* = \frac{\partial}{\partial (\omega/T)} 
 \frac{\sinh[(S+1/2)\omega/T]}{\sinh[\omega/2T]} \nonumber \\ 
&& \times \left[ \frac{\varepsilon}{m} \hat\omegav - 
 \frac{1}{m(\varepsilon+m)} (\hat\omegav \cdot {\bf p}) {\bf p} \right]
\eea
where ${\bf p}$ is the momentum and $\varepsilon$ the energy of the
particle in the frame where the fluid is rotating with a rigid
velocity field at a constant angular velocity $\omegav$, i.e. ${\bf v}
= \omegav \times {\bf x}$. It can be seen that the rest frame spin
vector has a component along its momentum, unlike in the
non-relativistic case, which vanishes in the low velocity limit
according to the non-relativistic formula (\ref{spin3}).
Note that eq.~(\ref{genspin}) is derived in the approximation 
$\omega/T \ll m/\varepsilon$~\cite{Becattini:2007nd} and the polarization is always 
less then unity.

The extension of these results to a fluid or a gas in a local
thermodynamic equilibrium situation, such as that which is assumed to
occur in the so-called hydrodynamic stage of the nuclear collision at
high energy, as well as the inclusion of quantum statistics effects,
requires more powerful theoretical tools. Particularly, if we want to describe 
the polarization of particles locally, a suitable approach requires the calculation
of the quantum-relativistic Wigner function and the spin tensor. By using such 
an approach, the mean spin vector of $1/2$ particles with four-momentum $p$, 
produced around point $x$ at the leading order in the thermal vorticity was 
found to be~\cite{Becattini:2013fla}:
\begin{equation}\label{eq:Pixp}
  S^\mu(x,p)= - \frac{1}{8m} (1-n_F) \epsilon^{\mu\rho\sigma\tau} p_\tau \varpi_{\rho\sigma}
\end{equation}
where $n_F = (1+\exp[\beta(x) \cdot p - \nu(x) Q/T(x)] +1)^{-1}$ is
the Fermi-Dirac distribution and $\varpi$ is given by eq.~\ref{thvort}
at the point $x$. This equation is suitable for the situation of
relativistic heavy ion collisions, where one deals with a local
thermodynamic equilibrium hypersurface $\Sigma$ where hydrodynamic
stage ceases and particle description sets in. It is the leading local thermodynamic 
equilibrium expression and it does not include dissipative corrections. It has 
been recovered with a different approach in ref.~\cite{Fang:2016vpj}. It is worth 
emphasizing that, according to the formula (\ref{eq:Pixp}), thermal vorticity 
rather than kinematical vorticity $\partial_\mu u_\nu - \partial_\nu u_\mu$ is 
responsible for the mean particle spin. There is a deep theoretical reason for this: the
four-vector $\beta$ in eq.~(\ref{thvort}) is a more fundamental vector for 
thermodynamic equilibrium in relativity than the velocity $u$ because it becomes 
a Killing vector field at global equilibrium \cite{Becattini:2012tc}. Hence, the 
expansion of the equilibrium, or local equilibrium, density operator, involves 
$\beta$ gradients as a parameter and not the gradients of velocity and temperature
separately \cite{betaframe}. To illustrate this statement, it is worth mentioning 
that, in a relativistic rotating gas at equilibrium, with velocity field 
${\bf v} = \omegav \times {\bf x}$ and $T = T_0/\sqrt{1-v^2}$, with $T_0$ constant, 
$\varpi$ is a constant, whereas the kinematical vorticity is not.

It is instructive to check that the eq.~(\ref{eq:Pixp}) yields, in the non-relativistic and global
equilibrium limit, the formulae obtained in the first part of this Section. First of all, 
at low momentum, in eq.~(\ref{eq:Pixp}) one can keep only the term corresponding to $\tau=0$
and $p_0 \simeq m$, so that $S^0 \simeq 0$ and:
\be 
  S^\mu(x,p) \simeq - \epsilon^{\mu\rho\sigma 0}\frac{1-n_F}{8} \varpi_{\rho\sigma}
\ee
Then, the condition of global equilibrium makes the thermal vorticity field constant and equal to the 
ratio of a constant angular velocity $\omegav$ and a constant temperature $T$ \cite{Becattini:2012tc}
that is:
\be
 - \frac{1}{2} \epsilon^{ijk0} \varpi_{jk} = \frac{1}{T_0} \omega^i
\ee
Finally, in the Boltzmann statistics limit $1 - n_F \simeq 1$ and one finally gets the spin 
3-vector as:
\be
 \bS(x,p) \simeq \frac{1}{4} \frac{\omegav}{T} 
\ee
which is the same result as in eq.~(\ref{eq:SpinOneHalf}). 

The formula (\ref{eq:Pixp}) has another interesting interpretation: the mean spin vector
is proportional to the axial thermal vorticity vector seen by the particle along its motion,
that is comoving. Indeed, an antisymmetric tensor can be decomposed into two spacelike vectors, one 
axial and one polar, seen by an observer with velocity $u$ (the subscript c stands for comoving):
\be
   \varpi_{\rm c}^\mu = -\frac{1}{2} \epsilon^{\mu\rho\sigma\tau} \varpi_{\rho\sigma} u_\tau
   \qquad \alpha_{\rm c}^\mu = \varpi^{\mu\nu} u_\nu
\ee
in much the same way as the electromagnetic field tensor $F_{\mu\nu}$ can be decomposed into
a comoving electric and magnetic field. Thus, the eq.~(\ref{eq:Pixp}) can be rewritten as:
\begin{equation}\label{eq:Pixp2}
  S^\mu(x,p)= \frac{1}{4} (1-n_F) \varpi_{\rm c}^\mu
\end{equation}
like in the non-relativistic case, provided that $\varpi_{\rm c}^\mu$ is the thermal vorticity axial
vector in the particle comoving frame.

To get the experimentally observable quantity, that is the spin vector of some particle 
species as a function of the four-momentum, one has to integrate the above expressions over 
the {\em particlization} hypersurface $\Sigma$:
\be\label{eq:Pip}
 S^\mu(p)= \frac{\int d\Sigma_\lambda p^\lambda f(x,p) S^\mu(x,p)}{\int d\Sigma_\lambda 
 p^\lambda f(x,p)}
\ee
The mean spin vector i.e. averaged over momentum, of some $S=1/2$ particle species, can be then 
expressed as:
\begin{equation}\label{eq:Pi}
  S^\mu =\frac{1}{N} \int\frac{\di^3 \p}{p^0} \int d\Sigma_\lambda p^\lambda n_F(x,p) S^\mu(x,p)
\end{equation}
where $N=\int\frac{\di^3\p}{p^0}\int d\Sigma_\lambda p^\lambda n_F(x,p)$ is the average number of 
particles produced at the particlization surface. One can also derive the expression of the spin 
vector in the rest frame from (\ref{eq:Pi}) taking into account Lorentz invariance of most of 
the factors in it:
\begin{align}
  S^{*\mu} = \frac{1}{N} \int \frac{\di^3 \p}{p^0} \int d\Sigma_\lambda p^\lambda n_F(x,p) 
   S^{*\mu}(x,p)
\end{align}
Looking at the eq.~(\ref{eq:Pixp2}), one would say that a measurement of the mean spin vector
provides an estimate of the {\em mean} comoving thermal vorticity axial vector.

As has been mentioned, the formula (\ref{eq:Pixp}) applies to spin $1/2$ particles. However, a very
plausible extension to higher spins can be obtained by comparing the global equilibrium 
expression (\ref{genspin}) for particles with spin $S$ in the Boltzmann statistics, with the
first-order expansion in the thermal vorticity for spin $1/2$ eq.~(\ref{eq:Pixp}). Taking
into account that the thermal vorticity should replace $\omega/T$ and the $\omega/T \ll 1$
expansion in eq.~(\ref{spin4}), one obtains, in the Boltzmann limit:
\begin{equation}\label{pixpgen}
   S^\mu(x,p) \simeq - \frac{1}{2m} \frac{S(S+1)}{3} \epsilon^{\mu\rho\sigma\tau} p_\tau 
   \varpi_{\rho\sigma}
\end{equation}
and the corresponding integrations over the hypersurface $\Sigma$ and momentum similar to 
eqs.~(\ref{eq:Pip}) and (\ref{eq:Pi}).

Finally, we would like to mention that the formula (\ref{pixpgen}) could be naturally extended
to include the electromagnetic field by simply replacing $\varpi_{\rho\sigma}$ with $\varpi_{\rho\sigma}
+ \mu F_{\rho\sigma}/S$, in agreement with the non-relativistic distribution in eq.~(\ref{thermrho}).
\be\label{pixpgen2}
   S^\mu(x,p) \simeq - \frac{1}{2m} \frac{S(S+1)}{3} \epsilon^{\mu\rho\sigma\tau} p_\tau 
  \left( \varpi_{\rho\sigma} - \frac{\mu}{S} F_{\rho\sigma} \right)
\ee
and, by using the comoving axial thermal vorticity vector and the comoving magnetic field:
\be\label{pixpgen3}
   S^\mu(x,p) \simeq \frac{S(S+1)}{3} \left( \varpi_{\rm c}^\mu + \frac{\mu}{S} B_{\rm c}^\mu \right)
\ee
%

\section{$\Lambda$ polarization measurement}\label{lambda}

The most straightforward way to detect a global polarization in relativistic nuclear collisions
is focussing on $\Lambda$ hyperons. As they decay weakly violating parity, in the $\Lambda$ 
rest frame the daughter proton is predominantly emitted along the $\Lambda$ polarization:
\be\label{eq:lambda}
\frac{dN}{d\Omega^*} =\frac{1}{4\pi} \left( 1+\alpha_\Lambda \bP^*_\Lambda \cdot \hat{\bp}^*
\right),
\ee
where $\alpha_\Lambda=-\alpha_\Lambdabar\approx0.642$ is the $\Lambda$ decay constant~\cite{Agashe:2014kda}.
$\hat{\bp}^*$ is the unit vector along the proton momentum and ${\bf P}^*$ the polarization
vector of the $\Lambda$, both in $\Lambda$'s rest frame.

For a global polarization measurement, one also needs to know the direction of the total angular 
momentum, along which the local thermal vorticity will be preferentially aligned. This direction 
can be reconstructed by measuring the directed flow of the projectile spectators (which 
conventionally is taken as a positive $x$ direction in the description of any anisotropic
flow~\cite{Voloshin:2008dg}). Recently it was shown that spectators, on average, deflect outward 
from the centerline of the collision~\cite{Voloshin:2016ppr}. Thus, measuring this deflection
provides information about the orientation of the nuclei during the collision (i.e. the impact 
parameter $\mathbf{b}$) and the direction of the angular momentum. One can also use for this purpose 
the flow of produced particles if their relative orientation with respect to the spectator flow 
is known. For heavy ion collisions the direction of the system orbital momentum on average
coincides with the direction of the magnetic field.

Finally, because the reaction plane angle can not be reconstructed exactly in experiments, one 
has to correct for the reaction plane resolution. In order to apply the standard flow methods 
for such a correction, it is convenient first to 'project' the distribution eq.\ref{eq:lambda} 
on the transverse plane, restricting the analysis to the difference in azimuths of the proton 
emission and that of the reaction plane. One arrives at~\cite{Abelev:2007zk}: 
\be
  P_\Lambda=\frac{8}{\pi \alpha_\Lambda}\frac{\mean{\sin(\psioep-\phi_p^*)}}{R^{(1)}_{\rm
    EP}}, 
\ee 
where $\psioep$ is the first harmonic (directed flow) event plane (e.g. determined by the 
deflection of projectile spectators) and $R^{(1)}_{\rm EP}$ is the corresponding event plane 
resolution (see Ref.~\cite{Abelev:2007zk} for the discussion of the detector acceptance effects).

It should be pointed out that in relativistic heavy ion collisions the electromagnetic field 
may also play a role in determining the polarization of produced particles. If we keep the
assumption of local thermodynamic equilibrium, one can apply the formulae (\ref{pixpgen2}),
(\ref{pixpgen3}). However, as yet, it is not clear if the spin degrees of freedom will 
respond to a variation of thermal vorticity as quickly as to a variation of the electromagnetic 
field. If the relaxation times were sizeably different, one would estimate thermal vorticity
and magnetic field from the measured polarization (see Section~\ref{feeddown}) at different 
times in the process. 
The magnetic moments of particles and antiparticles have opposite signs, so the 
effect of the electromagnetic field is a splitting in global polarization of particles and 
antiparticles. Particularly, the $\Lambda$ magnetic moment is $\mu_\Lambda\approx -0.61 \mu_N=-0.61 
e/(2m_p)$~\cite{Agashe:2014kda} and, under the assumption above, one can take advantage of a
difference in the polarization of primary $\Lambda$s and $\bar\Lambda$s (i.e. those emitted 
directly at hadronization) to estimate the (mean comoving) magnetic field:
\be \label{eq:Bsplitting}
 eB \approx -\Delta P^{\rm prim} m_p T/0.61
\ee
where $m_p$ is the  proton mass, and 
$\Delta P^{\rm prim}\equiv P^{\rm prim}_{\Lambda}-P^{\rm prim}_{\overline{\Lambda}}$ is the difference 
in polarization of {\it primary} $\Lambda$ and $\overline{\Lambda}$. An (absolute) difference 
in the polarization of primary $\Lambda$'s of of 0.1\% then would correspond to a magnetic field 
of the order of $ \sim 10^{-2} m_\pi^2$, well within the range of theoretical 
estimates~\cite{Tuchin:2014hza,Gursoy:2014aka,McLerran:2013hla}.
However, we warn that equation~\ref{eq:Bsplitting} should not be applied to experimental measurements 
without a detailed accounting for polarized feed-down effects, which are discussed in
Section~\ref{feeddown}.

Finally, we note that a small difference between $\Lambda$ and $\bar\Lambda$ polarization could also be 
due to the finite baryon chemical potential making the factor $(1-n_F)$ in eq.~(\ref{eq:Pixp})
different for particles and antiparticles; this Fermi statistics effect might be relevant only 
at low collision energies.

\section{Spin alignment of vector mesons}\label{align}

The global polarization of vector mesons, such as $\phi$ or $K^*$, can be accessed via the so-called spin
alignment~\cite{Liang:2004xn,Abelev:2008ag}. Parity is conserved in the strong decays of those particles  
and, as a consequence, the daughter particle distribution is the same for the states $S_z=\pm 1$. In fact, 
it is different for the state $S_z=0$, and this fact can be used to determine a polarization of the parent 
particle. By referring to eq.~(\ref{proba}), in the thermal approach the deviation of the probability for 
the state $S_z=0$ from 1/3, is only of the second order in $\tomega$: 
\be p_0=\frac{1}{1+2\cosh\tomega_{\rm c}}
\approx\frac{1}{3+\tomega_{\rm c}^2} \approx \frac{1}{3}(1-\tomega_{\rm c}^2/3) , 
\ee
which could make this measurement difficult. Similarly difficult will be the detection of the global 
polarization with the help of other strong decay channels, e.g. proposed in Ref.~\cite{Shuryak:2016hor}.

\section{Accounting for decays} \label{feeddown}

According to eq.~(\ref{pixpgen2})
(or, in the non-relativistic limit, equations~\ref{spin4}-\ref{eq:s3}), the polarization of 
primary $\Lambda$ hyperons provides a measurement 
of the (comoving) thermal vorticity and the (comoving) magnetic field of the system that emits them. However, 
only a fraction of all detected $\Lambda$ and $\bar{\Lambda}$ hyperons are produced directly at
the hadronization stage and are thus {\em primary}. Indeed, a large fraction thereof stems from decays 
of heavier particles and one should correct for feed-down from higher-lying resonances when trying
to extract information about the vorticity and the magnetic field from the measurement of polarization.
Particularly, the most important feed-down channels involve the strong decays of $\Sigma^*\rightarrow
\Lambda + \pi$, the electromagnetic decay $\Sigma^0\rightarrow \Lambda +\gamma$, and the weak decay $\Xi
\rightarrow \Lambda + \pi$.
\begin{table}
\vspace{10pt}
\begin{tabular}{|c|c|}
\hline
 Decay & $C$    \\ \hline
parity-conserving:~  $\sfrac{1}{2}^+ \to \sfrac{1}{2}^+ \;\; 0^-$    &   $-1/3$     \\ \hline
parity-conserving:~  $\sfrac{1}{2}^- \to \sfrac{1}{2}^+ \;\; 0^- $    &     1        \\ \hline
parity-conserving:~  $\sfrac{3}{2}^+ \to \sfrac{1}{2}^+ \;\; 0^- $   &   $1/3$      \\ \hline
parity-conserving:~  $\sfrac{3}{2}^- \to \sfrac{1}{2}^+ \;\; 0^- $    &   $-1/5$     \\ \hline
$\Xi^0\rightarrow\Lambda+\pi^0$ & +0.900 \\ \hline
$\Xi^-\rightarrow\Lambda+\pi^-$ & +0.927 \\ \hline
$\Sigma^0\rightarrow\Lambda+\gamma$ & -1/3 \\ \hline
\end{tabular}
\caption{Polarization transfer factors $C$ (see eq.~(\ref{linear})) for important decays 
  $X \to \Lambda(\Sigma) \pi$}
\label{ctable}
\end{table}

When polarized particles decay, their daughters are themselves polarized because of angular momentum 
conservation. The amount of polarization which is inherited by the daughter particle, or transferred from
the parent to the daughter, in general depends on the momentum of the daughter in the rest frame of the
parent. As long as one is interested in the {\em mean}, momentum-integrated, spin vector in the rest frame, 
a simple linear rule applies (see Appendix A), that is:
\be\label{linear}
  {\bf S}^*_D = C {\bf S}^*_P
\ee
where $P$ is the parent particle, $D$ the daughter and $C$ a coefficient whose expression (see Appendix A) 
may or may not depend on the dynamical amplitudes. In many two-body decays, the conservation laws constrain
the final state to such an extent that the coefficient $C$ is {\em independent} of the dynamical matrix
elements. This happens, e.g., in the strong decay $\Sigma^*(1385) \to \Lambda \pi$ and the electromagnetic
$\Sigma^0 \to \Lambda \gamma$ decay, whereas it does not in $\Xi \to \Lambda \pi$ decays, which is a
weak decay. 

If the decay products have small momenta compared to their masses, one would expect that the spin 
transfer coefficient $C$ was determined by the usual quantum-mechanical angular momentum addition rules 
and Clebsch-Gordan coefficients, as the spin vector would not change under a change of frame. Surprisingly, 
this holds in the relativistic case provided that the coefficient $C$ is independent of the dynamics,
as it is shown in Appendix A. In this case, $C$ is independent of Lorentz factors $\beta$ or $\gamma$
of the daughter particles in the rest frame of the parent, unlike naively expected. This feature makes $C$
a simple rational number in all cases where the conservation laws fully constrain it. 
The polarization transfer coefficients $C$ of several important baryons decaying to $\Lambda$s are 
reported in table~(\ref{ctable}) and their calculation described in detail in Appendix A.

Taking the feed-down into account, the measured mean $\Lambda$ spin vector along the angular momentum
direction can then be expressed as:
\be
 {\bf S}^{*,{\rm meas}}_{\Lambda} = 
 \sum_R \left[f_{\Lambda R} C_{\Lambda R} - \tfrac{1}{3}f_{\Sigma^0R}C_{\Sigma^0R}\right] {\bf S}^*_R .
\label{eq:decay}
\ee
This formula accounts for direct feed-down of a particle-resonance $R$ to a $\Lambda$, as well 
as the two-step decay $R\rightarrow\Sigma^0\rightarrow\Lambda$; these are the only significant 
feed-down paths to a $\Lambda$. In the eq.(~\ref{eq:decay}), $f_{\Lambda R}$ ($f_{\Sigma^0 R}$) 
is the fraction of measured $\Lambda$'s coming from $R\rightarrow\Lambda$ ($R\rightarrow\Sigma^0\rightarrow\Lambda$).
The spin transfer to the $\Lambda$ in the direct decay is denoted $C_{\Lambda R}$, while
$C_{\Sigma^0 R}$ represents the spin transfer from $R$ to the daughter $\Sigma^0$.
The explicit factor of $-\tfrac{1}{3}$ is the spin transfer coefficient from the $\Sigma^0$ 
to the daughter $\Lambda$ from the decay $\Sigma^0\rightarrow\Lambda+\gamma$.

In terms of polarization (see eq.~(\ref{spin4})):
\be\label{eq:decay2}
P^{\rm meas}_{\Lambda} = 2\sum_R \left[f_{\Lambda R} C_{\Lambda R} -\tfrac{1}{3}f_{\Sigma^0 R}C_{\Sigma^0 R}\right] S_R P_R 
\ee
where $S_R$ is the spin of the particle $R$. The sums in equations~(\ref{eq:decay}) and~(\ref{eq:decay2})
 are understood to include terms for the contribution of primary $\Lambda$s and $\Sigma^0$s.
These equations are readily extended to include additional multiple-step decay chains that terminate 
in a $\Lambda$ daughter, although such contributions would be very small.

Therefore, in the limit of small polarization, the polarizations of measured (including primary 
as well as secondary) $\Lambda$ and $\overline{\Lambda}$ are linearly related to the mean (comoving) 
thermal vorticity and magnetic field according to eq.~(\ref{pixpgen3}) or eq.~(\ref{spin4}), and 
these physical quantities may be extracted from measurement as:
\begin{widetext}
\begin{equation}\label{eq:FeeddownMatrix3}
\left(\begin{array}{c}\tomega_{\rm c} \\[25pt] B_{\rm c}/T\end{array}\right)
=
\left[\begin{array}{l l}
  \frac{2}{3}\sum\limits_{R}
    \left(f_{\Lambda R}\, C_{\Lambda R} -\frac{1}{3}f_{\Sigma^0 R}\,C_{\Sigma^0 R} \right)\, S_R (S_R+1)
   \quad  & \quad
  \frac{2}{3}\sum\limits_{R}
    \left(f_{\Lambda R}\, C_{\Lambda R} -\frac{1}{3}f_{\Sigma^0 R}\,C_{\Sigma^0 R} \right)\, (S_R+1)\, \mu_R
  \\[20pt]
  \frac{2}{3}\sum\limits_{\overline{R}}
     \left(f_{\overline{\Lambda} \overline{R}}\, C_{\overline{\Lambda} \overline{R}} - 
     \frac{1}{3} f_{\overline{\Sigma}^0 \overline{R}}\, C_{\overline{\Sigma}^0 \overline{R}}\right)\, S_{\overline{R}} (S_{\overline{R}}+1) 
   \quad  & \quad
  \frac{2}{3}\sum\limits_{\overline{R}}
    \left(f_{\overline{\Lambda} \overline{R}}\, C_{\overline{\Lambda} \overline{R}} - 
    \frac{1}{3} f_{\overline{\Sigma}^0 \overline{R}}\, C_{\overline{\Sigma}^0 \overline{R}}\right)\,(S_{\overline{R}}+1)\, \mu_{\overline{R}}
\end{array}\right]^{\mathlarger{-1}}
\left(\begin{array}{c}P^{\rm meas}_{\Lambda} \\[20pt] P^{\rm meas}_{\overline{\Lambda}}\\[5pt]\end{array}\right)  . 
\end{equation}
\end{widetext}
In the eq.~(\ref{eq:FeeddownMatrix3}), $\overline{R}$ stands for antibaryons that feed down into measured 
$\overline{\Lambda}$s. The polarization transfer is the same for baryons and antibaryons 
($C_{\overline{\Lambda R}}=C_{\Lambda R}$) and the magnetic moment has opposite sign ($\mu_{\overline{R}}=-\mu_{R}$).

According to the THERMUS model~\cite{Wheaton:2004qb}, tuned to reproduce semi-central Au+Au collisions at 
$\sqrt{s_{\rm NN}}=19.6$~GeV, fewer than 25\% of measured $\Lambda$s and $\overline{\Lambda}$s are primary, 
while more than 60\% may be attributed to feed-down from primary $\Sigma^*$, $\Sigma^0$ and $\Xi$ baryons.

The remaining $\sim15\%$ come from small contributions from a large number higher-lying resonances such as 
$\Lambda(1405),\Lambda(1520),\Lambda(1600),\Sigma(1660)$ and $\Sigma(1670)$. We find that, for $B=0$, their 
contributions to the measured $\Lambda$ polarization largely cancel each other, due to alternating signs of 
the polarization transfer factors. Their net effect, then, is essentially a 15\% ``dilution,''
contributing $\Lambda$s to the measurement with no effective polarization.
Since the magnetic moments of these baryons are unmeasured, it is not clear what their contribution to 
$P_{\Lambda {\rm meas}}$ would be when $B\neq0$.  However, it is reasonable to assume it would be small, 
as the signs of both the transfer coefficients and the magnetic moments will fluctuate.

Accounting for feed-down is crucial for quantitative estimates of vorticity and magnetic field based on 
experimental measurements of the global polarization of hyperons, as we illustrate with an example, using 
$\sqrt{s_{\rm NN}}=19.6$~GeV THERMUS feed-down probabilities. Let us assume that the thermal vorticity is 
$\tomega=0.1$ and the magnetic field is $B=0$. In this case, according to eq.~(\ref{eq:SpinOneHalf}), 
the primary hyperon polarizations are $P^{\rm prim}_{\Lambda}=P^{\rm prim}_{\overline{\Lambda}}=0.05$.
However, the measured polarizations would be $P^{\rm meas}_{\Lambda}=0.0395$ and $P^{\rm meas}_{\overline{\Lambda}}=0.0383$.
The two measured values differ because the finite baryochemical potential at these energies leads to 
slightly different feed-down fractions for baryons and anti-baryons. 

Hence, failing to account for feed-down when using equation~\ref{eq:SpinOneHalf} would lead to a $\sim20\%$ 
underestimate of the thermal vorticity. Even more importantly, if the splitting between $\Lambda$ and 
$\overline{\Lambda}$ polarizations were attributed entirely to magnetic effects (i.e. if one neglected 
to account for feed-down effects), equation~(\ref{eq:Bsplitting}) would yield an erroneous estimate 
$B\approx-0.015m_\pi^2$. This erroneous estimate has roughly the magnitude of the magnetic field expected 
in heavy ion collisions, but points the in the ``wrong'' direction, i.e. opposite the vorticity.
In other words, in the absence of feed-down effects, a magnetic field is expected to cause 
$P_{\overline{\Lambda}}>P_{\Lambda}$, whereas feed-down in the absence of a magnetic field will produce 
a splitting of the opposite sign.

\section{Summary and conclusions}\label{conclu}

The nearly-perfect fluid generated in non-central heavy ion collisions is characterized by a huge 
vorticity and magnetic field, both of which can induce a global polarization of the final
hadrons. Conversely, a measurement of polarization makes it possible
to estimate the thermal vorticity as well as the electromagnetic field developed in the plasma 
stage of the collision. As the thermal vorticity appears to be strongly dependent on the 
hydrodynamic initial conditions, polarization is a very sensitive probe of the QGP formation
process. Pinning down (thermal) vorticity and magnetic field is also very important for the 
quantitative assessment of thus-far unobserved QCD effects, such as the chiral magnetic and 
chiral vortical effects. 

We have summarized and elucidated the thermal approach to the calculation of the polarization
of particles in relativistic heavy ion collisions, based on the assumption of local thermodynamic
equilibrium of the spin degrees of freedom at hadronization. We have put forward an 
extension of the formulae for spin $\sfrac{1}{2}$  particles to particles with any 
spin, with an educated guess based on the global equilibrium case. The extension to 
any spin is needed to account for feed-down contributions that are crucial to make 
a proper estimate of the polarization at the hadronization stage. 

We have discussed in detail how polarization is transferred to the decay products in a
decay process and shown that a simple linear propagation rule applies to the momentum-averaged 
rest-frame spin vectors. We have developed the general formulae for the polarization transfer 
coefficients in two-body decays and carried out the explicit calculations for the most important 
decays involving a $\Lambda$ hyperon. We have shown how to take the decays into account for 
the extraction of thermal vorticity and magnetic field. It should be stressed, though, 
that the extraction of such quantities at hadronization relies on the aforementioned assumption 
of local thermodynamic equilibrium; it is still unclear whether this is correct for the 
electromagnetic field term.

The feed-down corrections can be significant, reducing the measured polarizations by $\sim20\%$, 
as compared to the polarization of primary particles at RHIC energies.
More importantly, feed-down may generate a splitting between measured $\Lambda$ and $\overline{\Lambda}$
polarizations of roughly the same magnitude as the splitting expected from magnetic effects.  
Fortunately, at finite baryochemical potential, the two splittings have opposite sign, so that
feed-down effects should not ``artificially'' mock up magnetic effects.

Finally, it must be pointed out that there is a further effect, in fact much harder to assess,
which can affect the reconstruction of the polarization of primary particles, that is 
post-hadronization interactions. Indeed, hadronic elastic interaction may involve a spin 
flip which, presumably, randomizes the spin direction of primary as well as secondary particles, 
thus decreasing the estimated mean global polarization.

\section*{Acknowledgments}

We would like to thank Bill Llope for providing feed-down estimates based on the THERMUS model.

This material is based upon work supported in part by the U.S. Department of Energy Office of Science, Office of Nuclear Physics under Award Number DE-FG02-92ER-40713; the U.S. National Science Foundation under Award Number 1614835; the University of Florence grant {\it Fisica dei plasmi relativistici: teoria e applicazioni moderne} and the INFN project {\it Strongly Interacting Matter}.

\appendix

\section{Polarization transfer in two-body decay}
  
We want to calculate the polarization which is inherited by the $\Lambda$ hyperons in decays of 
polarized higher lying states and, particularly, $\Sigma^* \rightarrow \Lambda \pi$, $\Sigma^0 
\rightarrow \Lambda \gamma$ and $\Xi \rightarrow \Lambda \pi$. The goal is to determine 
the mean spin vector in the $\Lambda$ rest frame, as a function of the mean spin vector of 
the decaying particle in {\em its} rest frame. We will finally show that the equation 
(\ref{linear}) applies and we will determine the exact expression of the coefficient $C$.

We will work out the exact relativistic result. In a relativistic framework, the use of the 
helicity basis is very convenient; for a complete description of the helicity and alternative 
spin formalisms, we refer the reader to refs.~\cite{moussa,wukitung,chung} For a particle 
with spin $J$ and spin projection along the $z$ 
axis $M$ in its rest frame (in the rest frame helicity coincides with the eigenvalue of the spin 
operators $\widehat{S}$, conventionally $\widehat{S}_3$, see text) decaying into two particles 
$A$ and $B$, the final state $\ket{\psi}$ can be written as a superposition of states with 
definite momentum and helicities:
\begin{equation}
 \ket{\psi} \propto \sum_{\lambda_A,\lambda_B} \int \di \Omega \; D^J(\varphi,\theta,0)^{M*}_{\lambda}
  \ket{{\bf p},\lambda_A,\lambda_B} T^J(\lambda_A,\lambda_B) 
\end{equation}  
where ${\bf p}$ is the momentum of either particle, $\theta$ and $\varphi$ its spherical coordinates
$\di \Omega = \sin\theta \di \theta \di \varphi$ the corresponding infinitesimal solid angle, $D^J$ is 
the Wigner rotation matrix in the representation of spin $J$, $T^J(\lambda_A,\lambda_B)$ are the 
reduced dynamical amplitudes depending only on the final helicities and:
$$
  \lambda = \lambda_A - \lambda_B
$$

The mean relativistic spin vector of, e.g., the particle $A$ after the decay is given by:
$$
  S^\mu = \bra{\psi} \widehat{S}_A^\mu \ket{\psi}
$$
with $\braket{\psi}{\psi}=1$, hence:
\bea\label{spina1}
 S^\mu &=&\sum_{\lambda_A,\lambda_B,\lambda'_A} \int \di \Omega \; 
  D^J(\varphi,\theta,0)^{M*}_{\lambda}D^J(\varphi,\theta,0)^{M}_{\lambda'} \nonumber \\
  &\times& \bra{\lambda'_A} \widehat{S}_A^\mu \ket{\lambda_A} T^J(\lambda_A,\lambda_B)
  T^J(\lambda'_A,\lambda_B)^* \nonumber \\ 
  &\times& \left( {\sum_{\lambda_A,\lambda_B} \int \di \Omega \; 
  | D^J(\varphi,\theta,0)^{M*}_{\lambda} |^2 | T^J(\lambda_A,\lambda_B) |^2} \right)^{-1} \nonumber \\
  &=&\sum_{\lambda_A,\lambda_B,\lambda'_A} \int \di \Omega \; 
  D^J(\varphi,\theta,0)^{M*}_{\lambda}D^J(\varphi,\theta,0)^{M}_{\lambda'} \nonumber \\
  &\times& \bra{\lambda'_A} \widehat{S}_A^\mu \ket{\lambda_A} T^J(\lambda_A,\lambda_B)
  T^J(\lambda'_A,\lambda_B)^* \nonumber \\ 
  &\times& \left( \frac{4\pi}{2J+1}{\sum_{\lambda_A,\lambda_B}| T^J(\lambda_A,\lambda_B) |^2} 
  \right)^{-1}
\eea
where we have used the known integrals of the Wigner $D$ matrices and the fact that the operator 
$\widehat{S}_A$ does not change the momentum eigenvalues as well as the helicity of the particle 
$B$. This operator can be decomposed as in eq.~(\ref{spindec}), with $n_i(p)$ being three spacelike 
unit vectors orthogonal to the four-momentum $p$. They can be obtained by applying the so-called 
{\em standard Lorentz transformation} $[p]$ turning the unit time vector $\hat t$ into the direction of 
the four-momentum $p$ \cite{moussa}, to the three space axis vectors ${\bf e}_i$, namely:
$$
  n_i(p) = [p]({\bf e}_i)
$$  
so that (we have temporarily dropped the subscript $A$ for the sake of simplicity):
\be\label{spina2}
 \widehat{S} = \sum_i \widehat{S}_i n_i(p) = [p](\sum_i \widehat{S}_i {\bf e}_i)
\ee
by taking advantage of the linearity of $[p]$. It is convenient to rewrite the sum in the argument
of $[p]$ along the spherical vector basis $e_+,e_-,e_0$ which is used to define the $D^J$ matrix
elements:
\begin{eqnarray*}
  {\bf e}_+ &=& -\frac{1}{\sqrt{2}}({\bf e}_1 + \ii {\bf e}_2) \\
  {\bf e}_- &=& \frac{1}{\sqrt{2}}({\bf e}_1 - \ii {\bf e}_2) \\
  {\bf e}_0 &=& {\bf e}_3
\end{eqnarray*}
so that:
\be\label{spina3}
  \sum_i \widehat{S}_i {\bf e}_i = - \frac{1}{\sqrt{2}} \widehat{S}_- {\bf e}_+ + 
  \frac{1}{\sqrt{2}} \widehat{S}_+ {\bf e}_- + \widehat{S}_0 {\bf e}_0  
\ee  
where $\widehat{S}_\pm = \widehat{S}_1 \pm i \widehat{S}_2$ are the familiar ladder operators. With 
these operators, we can now easily calculate the spin matrix elements in eq.~(\ref{spina1}) because 
their action onto helicity kets $\ket{\lambda}$ is precisely the familiar one onto eigenstates of 
the $z$ projection of angular momentum with eigenvalue $\lambda$, e.g.:
$$
 \bra{\lambda'} \widehat{S}_0 \ket{\lambda} = \lambda \delta_{\lambda,\lambda'}
$$
and in general, using eqs.~(\ref{spina2}) and (\ref{spina3}), we can write:
\be\label{spina4}
 \bra{\lambda'_A} \widehat{S}_A \ket{\lambda_A} = \sum_{n=-1}^1 a_n
 \bra{\lambda'_A} \widehat{S}_{A,-n} \ket{\lambda_A} [p]({\bf e}_n) 
\ee 
where $a_n = -n/\sqrt{2} + \delta_{n,0}$.

To work out the eq.~(\ref{spina4}), we need to find an expression of the standard transformation $[p]$.
In principle, it can be freely chosen, but the choice which makes $\lambda$ the particle
helicity \cite{weinberg,chung} is the composition of a Lorentz boost along the $z$ axis of hyperbolic 
angle $\xi$ such that $\sinh \xi = \|{\bf p}\|/m$, followed by a rotation around the $y$ axis of angle $\theta$ 
and a rotation around the $z$ axis by an angle $\varphi$ (see above):
$$
  [p] = {\sf R}_z(\varphi) {\sf R}_y(\theta) {\sf L}_z(\xi)
$$
Thus:
\begin{eqnarray*}
 [p]({\bf e}_\pm) &=& {\sf R}_z(\varphi) {\sf R}_y(\theta) {\sf L}_z(\xi) ({\bf e}_\pm) \nonumber \\
 &=& {\sf R}_z(\varphi) {\sf R}_y(\theta)({\bf e}_\pm) = \sum_{l=-1}^1 D^1(\varphi,\theta,0)^l_{\pm 1} {\bf e}_l
\end{eqnarray*}
because $e_\pm$ is invariant under a boost along the $z$ axis. Conversely, ${\bf e}_0$ is not invariant under 
the Lorentz boost and:
\begin{eqnarray*}
 [p]({\bf e}_0) &=& \cosh \xi {\sf R}_z(\varphi) {\sf R}_y(\theta)({\bf e}_0) + \sinh \xi {\sf R}_z(\varphi) 
 {\sf R}_y(\theta)(\hat t) \nonumber \\
 &=& \sum_{l=-1}^1 \frac{\varepsilon}{m} D^1(\varphi,\theta,0)^l_0 {\bf e}_l + \frac{\rm p}{m} \hat t 
\end{eqnarray*}
where ${\rm p})= \|{\bf p}\|$, $\varepsilon = \sqrt{{\rm p}^2+m^2}$ is the energy and $\hat t$ is the 
unit vector in the time direction. We can now plug the above two equations into the eq.~(\ref{spina4}) to get:
\bea\label{spina5}
&& \bra{\lambda'_A} \widehat{S}_A \ket{\lambda_A} = \sum_{l,n} b_n
 D^1(\varphi,\theta,0)^l_n \bra{\lambda'_A} \widehat{S}_{A,-n} \ket{\lambda_A} {\bf e}_l 
 \nonumber \\
&& + \lambda_A \delta_{\lambda_A,\lambda'_A} \frac{\rm p}{m} \hat t
\eea 
where $b_n = -n/\sqrt{2} + \gamma \delta_{n,0}$ with $\gamma=\varepsilon/m$ the Lorentz factor of
the decayed particle $A$ in the rest frame of the decaying particle.

We can now write down the fully expanded expression of the mean spin vector $S$
in eq.~(\ref{spina1}). The time component is especially simple; by using the eq.~(\ref{spina5}) 
one has:
\bea\label{timecomp}
  S^0 &=& \frac{\rm p}{m} \sum_{\lambda_A,\lambda_B} \lambda_A \int \di \Omega \; 
  |D^J(\varphi,\theta,0)^{M*}_{\lambda}|^2 |T^J(\lambda_A,\lambda_B)|^2
  \nonumber \\ 
  &\times& \left( \frac{4\pi}{2J+1}{\sum_{\lambda_A,\lambda_B}| T^J(\lambda_A,\lambda_B) |^2} 
  \right)^{-1} 
\eea  
and after integrating over $\Omega$:
\be\label{timecomp2}
 S^0 = \frac{\rm p}{m} \frac{\sum_{\lambda_A,\lambda_B} 
  \lambda_A |T^J(\lambda_A,\lambda_B) |^2}
  {\sum_{\lambda_A,\lambda_B}| T^J(\lambda_A,\lambda_B) |^2} 
\ee  
Similarly, the space component reads:
\bea\label{spacecomp}
 && {\bf S} = \!\!\!\! \sum_{\lambda_A,\lambda_B,\lambda'_A} \!\!\! T^J(\lambda_A,\lambda_B) 
  T^J(\lambda'_A,\lambda_B)^* \sum_{n,l} \bra{\lambda'_A} \widehat{S}_{A,-n} \ket{\lambda_A}  \nonumber \\
 && \times b_n \int \di \Omega \; D^J(\varphi,\theta,0)^{M*}_{\lambda} D^J(\varphi,\theta,0)^{M}_{\lambda'}   
  D^1(\varphi,\theta,0)^l_{n} {\bf e}_l \nonumber \\
 && \times \left( \frac{4\pi}{2J+1}{\sum_{\lambda_A,\lambda_B}| T^J(\lambda_A,\lambda_B) |^2} \right)^{-1}
\eea

We note that the integrands in the angular variables $\theta,\varphi$ in both eqs.~(\ref{timecomp}) and
(\ref{spacecomp}) are proportional to the mean relativistic spin vector at some momentum ${\bf p}$, 
that is $S(p)$. The angular integrals in the eq.~(\ref{spacecomp}) are known and can be expressed 
in terms of Clebsch-Gordan coefficients:
\bea\label{spacecomp2}
 && {\bf S} = \!\!\!\! \sum_{\lambda_A,\lambda_B,\lambda'_A} \!\!\! T^J(\lambda_A,\lambda_B) 
  T^J(\lambda'_A,\lambda_B)^* \sum_{n,l} \bra{\lambda'_A} \widehat{S}_{A,-n} \ket{\lambda_A}  \nonumber \\
 && \!\!\!\times b_n \bra{JM} J1 \ket{Ml} \bra{J \lambda} J1 \ket{\lambda'n}  {\bf e}_l 
  \left({\sum_{\lambda_A,\lambda_B}| T^J(\lambda_A,\lambda_B) |^2} \right)^{-1} \nonumber \\
 && \!\!\!= \sum_{\lambda_A,\lambda_B,\lambda'_A} \!\!\! T^J(\lambda_A,\lambda_B) 
  T^J(\lambda'_A,\lambda_B)^* \sum_{n} \bra{\lambda'_A} \widehat{S}_{A,-n} \ket{\lambda_A}  \nonumber \\
 && \!\!\! \times b_n \bra{JM} J1 \ket{M0} \bra{J \lambda} J1 \ket{\lambda'n}  {\bf e}_0 
  \left({\sum_{\lambda_A,\lambda_B}| T^J(\lambda_A,\lambda_B) |^2} \right)^{-1} \nonumber \\
\eea
Note that the only non-vanishing spatial component of the mean relativistic spin vector is the 
along the $z$ axis, being proportional to ${\bf e}_0 = {\bf e}_3$. This is a result of rotational 
invariance, as the decaying particle is polarized along this axis by construction. 

What we have calculated so far is the mean relativistic spin vector in the decaying particle rest
frame. However, one is also interested in the same vector in the {\em decayed} (that is $A$) particle rest 
frame. For some momentum ${\bf p}$, it can be obtained by means of a Lorentz boost:
$$
 {\bf S}^*(p) = {\bf S}(p) - \frac{{\bf p}}{\varepsilon(\varepsilon+m)} {\bf S}(p) \cdot {\bf p}
$$
Since ${\bf S}(p) \cdot {\bf p} = S^0(p) \varepsilon$ as $S$ is a four-vector orthogonal to $p$, we can 
obtain the mean, i.e. momentum integrated, vector: 
\bea\label{ptransf}
 && {\bf S}^* = \langle {\bf S}^*(p) \rangle = \langle {\bf S}(p) \rangle  - \frac{1}{\varepsilon+m} 
  \langle {\bf p} S^0(p)\rangle \nonumber \\
 && = {\bf S} - \frac{1}{\varepsilon+m} \langle {\bf p} S^0(p)\rangle
\eea
The first term on the right hand side is the vector in eq.~(\ref{spacecomp2}), while for the second term we
have, from eq.~(\ref{timecomp}) and using:
$$
  {\bf p} = {\rm p} \sum_{l=-1}^1 D^1(\varphi,\theta,0)^l_0 {\bf e}_l
$$
\bea\label{boostcomp}
 && \langle {\bf p} S^0(p) \rangle = \frac{{\rm p}^2}{m} \sum_{\lambda_A,\lambda_B} \lambda_A 
  |T^J(\lambda_A,\lambda_B)|^2 \sum_{l=-1}^1 {\bf e}_l \nonumber \\
 && \times \int \di \Omega \; |D^J(\varphi,\theta,0)^{M*}_{\lambda}|^2 D^1(\varphi,\theta,0)^l_0 \nonumber \\ 
 && \times \left( \frac{4\pi}{2J+1}{\sum_{\lambda_A,\lambda_B}| T^J(\lambda_A,\lambda_B) |^2} 
  \right)^{-1} \nonumber \\
 && = \frac{{\rm p}^2}{m} \sum_{\lambda_A,\lambda_B} \lambda_A |T^J(\lambda_A,\lambda_B)|^2 \sum_{l=-1}^1 {\bf e}_l \nonumber \\ 
 && \times \bra{JM} J1 \ket{M l} \bra{J\lambda} J1 \ket{\lambda 0}
  \left({\sum_{\lambda_A,\lambda_B}| T^J(\lambda_A,\lambda_B) |^2} \right)^{-1} \nonumber \\
 && = \frac{{\rm p}^2}{m} \sum_{\lambda_A,\lambda_B} \lambda_A |T^J(\lambda_A,\lambda_B)|^2 
  \bra{JM} J1 \ket{M 0} \bra{J\lambda} J1 \ket{\lambda 0} {\bf e}_0 \nonumber \\
 && \times \left({\sum_{\lambda_A,\lambda_B}| T^J(\lambda_A,\lambda_B) |^2} \right)^{-1}
\eea
By substituting eqs.~(\ref{boostcomp}) and (\ref{spacecomp2}) into the eq.~(\ref{ptransf}) one finally gets:
\bea\label{proper}
 &&  {\bf S}^* = \sum_{\lambda_A,\lambda_B,\lambda'_A} \!\!\! T^J(\lambda_A,\lambda_B) 
  T^J(\lambda'_A,\lambda_B)^* \sum_{n} \bra{\lambda'_A} \widehat{S}_{A,-n} \ket{\lambda_A}  \nonumber \\
 && \times c_n \bra{JM} J1 \ket{M0} \bra{J \lambda} J1 \ket{\lambda'n}  {\bf e}_0 
  \left({\sum_{\lambda_A,\lambda_B}| T^J(\lambda_A,\lambda_B) |^2} \right)^{-1} \nonumber \\
\eea
with:
\be\label{cn}
 c_n = - \frac{n}{\sqrt{2}} + \left( \gamma - \frac{\beta^2\gamma^2}{\gamma+1} \right) \delta_{n,0}
 = - \frac{n}{\sqrt{2}} + \delta_{n,0}
\ee
Note the disappearance of any dependence on the energy of the decay product, i.e. on the masses 
involved in the decay, once the mean relativistic spin vector is back-boosted to its rest frame 
(see also eqs~(\ref{general}),(\ref{coeffic}). 

The mean spin vector in eq.~(\ref{proper}) pertains to a decaying particle in the state $\ket{JM}$, that
is in a definite eigenstate of its spin operator $\widehat{S}_z$ in its rest frame. For a mixed state 
with probabilities $P_M$, one is to calculate the weighted average. Since:
$$
 \bra{JM} J1 \ket{M0} = \frac{M}{\sqrt{J(J+1)}} 
$$
the weighted average turns out to be:
\bea\label{polariz1}
 && {\bf S}^* = \sum_M M P_M {\bf e}_0 \sum_{\lambda_A,\lambda_B,\lambda'_A} \!\!\! 
  T^J(\lambda_A,\lambda_B) T^J(\lambda'_A,\lambda_B)^* \nonumber \\
 && \times \sum_{n=-1}^1 \bra{\lambda'_A} \widehat{S}_{A,-n} \ket{\lambda_A} 
  \frac{c_n}{\sqrt{J(J+1)}} \bra{J \lambda} J1 \ket{\lambda'n} \nonumber \\
 && \left({\sum_{\lambda_A,\lambda_B}| T^J(\lambda_A,\lambda_B) |^2} \right)^{-1}
\eea
Now, since $\sum_M M P_M {\bf e}_0$ is but the mean relativistic spin vector of the decaying particle,
from eq.~(\ref{polariz1}) we finally obtain that the mean spin vector of the decay product $A$ in
its rest frame is proportional to the spin vector of the decaying particle in its rest frame (see
eq.~(\ref{linear}):
\be\label{general}
  {\bf S}^*_A = C {\bf S}^*
\ee
with 
\bea\label{coeffic}
 && \!\!\!\!\!\! C = \!\!\!\! \sum_{\lambda_A,\lambda_B,\lambda'_A} \!\!\! T^J(\lambda_A,\lambda_B) 
  T^J(\lambda'_A,\lambda_B)^* \sum_{n=-1}^1 \bra{\lambda'_A} \widehat{S}_{A,-n} \ket{\lambda_A} \nonumber \\
 && \times \frac{c_n}{\sqrt{J(J+1)}} \bra{J \lambda} J1 \ket{\lambda'n}
  \left({\sum_{\lambda_A,\lambda_B}| T^J(\lambda_A,\lambda_B) |^2} \right)^{-1} \nonumber \\
\eea
Note that the Clebsch-Gordan coefficients involved in (\ref{coeffic}) can be written as:
\bea\label{cgc}
\!\!\!\!\!\!\!\!\!\!\!\!  
&& \bra{J \lambda} J1 \ket{\lambda 0} = \frac{\lambda}{\sqrt{J(J+1)}} \nonumber \\
\!\!\!\!\!\!\!\!\!\!\!\!  
&& \bra{J \lambda} J1 \ket{(\lambda\mp 1) \pm 1} = \mp \sqrt{\frac{(J\mp\lambda +1)(J\pm\lambda)}{2J(J+1)}} 
\eea
The proportionality between the two vectors as expressed by the eq.~(\ref{general}) could have been 
predicted as, once the momentum integration is carried out, the only possible direction of the mean 
spin vector of the decay product is the direction of the mean spin of the decaying particle. In fact,
the somewhat surprising feature of eq.~(\ref{coeffic}) is, as has been mentioned, the absence of an 
explicit dependence of $C$ on the masses involved in the decays as $c_n$ in eq.~(\ref{cn}) is independent of
them. There is of course an implicit dependence on the masses in the amplitudes $T^J$, but this can 
cancel out in several important instances.
  
If the interaction driving the decay is parity-conserving - what is the case for decays involving the 
strong and electromagnetic forces $\Sigma^* \rightarrow \Lambda \pi$ and $\Sigma^0 \rightarrow \Lambda \gamma$
- then there is a relation between the amplitudes \cite{chung}:
\be\label{phase}
  T^J(-\lambda_A,-\lambda_B) = \eta \eta_A \eta_B (-1)^{J-S_A-S_B} 
  \times T^J(\lambda_A,\lambda_B)
\ee 
where $\eta$ is the intrinsic parity of the decaying particle and $\eta_A,\eta_B$ those of
the {\em massive} decay products and $S_A,S_B$ their spins. A similar relation holds with $S=|\lambda|$ 
in eq.~(\ref{phase}) \cite{wukitung} if the particle is massless. Thus, in all cases, one has:
\be\label{phase2} 
  |T^J(-\lambda_A,-\lambda_B)|^2 = |T^J(\lambda_A,\lambda_B)|^2
\ee
The equations (\ref{phase}),(\ref{phase2}) have interesting consequences. First of all, from eq.~(\ref{timecomp2})
it can be readily realized that the time component of the mean relativistic spin vector vanishes. 
Secondly, if, because of the (\ref{phase}), only one independent reduced matrix element is left in 
eq.~(\ref{coeffic}), the final mean spin vector will be independent of the dynamics and determined 
only by the conservation laws. We will see that this is precisely the case for $\Sigma^* \rightarrow \Lambda \pi$ 
and $\Sigma^0 \rightarrow \Lambda \gamma$.

\subsection{$\Sigma^* \rightarrow \Lambda \pi$}  
  
In this case $\lambda_B = 0$, $\lambda =\lambda_A$, $J=3/2$ and $T^J(\lambda)$ is proportional to
$T^J(-\lambda)$ through a phase factor, which turns out to be $1$ from eq.~(\ref{phase}). Since 
$|\lambda|=1/2$ there is only one independent reduced helicity amplitude and so the coefficient
$C$ simplifies to:
\be\label{sigma1}
 C = \sum_{\lambda,\lambda'} \sum_{n=-1}^1 \bra{\lambda'} \widehat{S}_{A,-n} \ket{\lambda} 
  \frac{c_n}{\sqrt{J(J+1)}} \frac{\bra{J \lambda} J1 \ket{\lambda'n}}{2S_\Lambda + 1}
\ee
The three terms in the above sum with $n=-1,0,+1$ have to be calculated separately. For $n=0$ one 
obtains:
$$
 \frac{1}{2} \sum_{\lambda_A} \lambda^2_A \frac{1}{J(J+1)} = \frac{1}{15}
$$
where we have used the first equation in~(\ref{cgc}).

For $n=1$, the operator in eq.~(\ref{sigma1}) is $\widehat{S}_-$, which selects $\lambda'=-1/2$ 
and, correspondingly, $\lambda=1/2$. Similarly, for $n=-1$, the ladder operator in eq.~(\ref{sigma1})
selects the converse combination. From eq.~(\ref{cgc}), the Clebsch-Gordan coefficients turn out 
to have the same magnitude with opposite sign and, by using the eq.~(\ref{cn}), the contribution
of the $n=\pm1$ turns out to be the same, that is:
$$
 \frac{1}{2} \sqrt{\frac{8}{15}} \frac{1}{\sqrt{2}} \frac{1}{\sqrt{J(J+1)}} = \frac{2}{15}
$$
Therefore, the coefficient $C$ is:
\be\label{sigma2}
  C = \frac{1}{15} + 2 \frac{2}{15} = \frac{1}{3}
\ee
%

\subsection{$\Sigma^0 \rightarrow \Lambda \gamma$}  
 
This case is fully relativistic as one of the final particles is a photon, hence the helicity basis 
is compelling. Looking at the equation (\ref{spina1}) it can be seen that, for $J=1/2$:
$$
 |\lambda| = \lambda_A - \lambda_B = 1/2
$$
Since $B$ is a photon $\lambda_B = \pm 1$ and there are two cases:
\begin{eqnarray*}
 && \lambda_B =  1 \implies \lambda_A = 1/2  \implies \lambda = -1/2 \nonumber \\
 && \lambda_B = -1 \implies \lambda_A = -1/2  \implies \lambda = 1/2
\end{eqnarray*}
which in turn implies $\lambda = -\lambda_A$ and $\lambda_B = 2\lambda_A$ in eq.~(\ref{coeffic}). 
The same argument applies to $\lambda' = \lambda'_A -\lambda_B$, so we conclude that $\lambda_B
= 2 \lambda'_A$, whence $\lambda'_A = \lambda_A$ and $\lambda = \lambda'$. This in turn implies 
$n=0$ in the eq.~(\ref{coeffic}), which then reads, with $\lambda_B = 2\lambda_A$:
\bea\label{sigmaz1}
 && C = \sum_{\lambda_A} \lambda_A |T^J(\lambda_A,\lambda_B)|^2  \frac{1}{\sqrt{J(J+1)}} \nonumber \\
 && \times \bra{J -\lambda_A} J1 \ket{-\lambda_A 0} 
 \left({\sum_{\lambda_A} | T^J(\lambda_A,\lambda_B) |^2} \right)^{-1}
\eea
Like in the previous case, because of (\ref{phase2}), there is only one independent dynamical 
reduced squared matrix element, so eq.~(\ref{sigmaz1}) becomes:
\be\label{sigmaz2}
 C = \sum_{\lambda_A} \lambda_A \frac{(-\lambda_A)}{J(J+1)}\frac{1}{2 S_\Lambda +1}
\ee
where we have used the first equation in (\ref{cgc}). Replacing $J,S_\Lambda = 1/2$, we recover
the known result~\cite{Sucher:1965,Armenteros:1970eg}:
$$
  C = - \frac{1}{3} 
$$
%

\subsection{Other parity-conserving (strong and electromagnetic) decays}  
 
By using the same procedure as for the decay of $\Sigma^*$ it is possible to determine the factor
$C$ for more kinds of strong and electromagnetic decays into a $\sfrac{1}{2}^+$, such as $\Lambda$ or 
$\Sigma^0$ and a pion. The factors are reported in table~(\ref{ctable}).

\subsection{$\Xi \rightarrow \Lambda \pi$}  
 
This decay is weak, thus parity is not conserved and we cannot use the previous arguments. The 
polarization transfer in this decay has been studied in detail in the past, however, and the Lee-Yang 
formula for weak $\Xi$ decay quantifies the polarization of the daughter $\Lambda$ in terms of 
three parameters, $\alpha_{\Xi}$, $\beta_{\Xi}$, and $\gamma_{\Xi}$~\cite{Luk:2000zw,Huang:2004jp}:
\begin{widetext}
\begin{equation}\label{poladetail}
{\bf P}^*_{\Lambda}=\frac{\left(\alpha_{\Xi}+{\bf P}^*_{\Xi}\cdot \hat{\bf p}_\Lambda\right)\hat{\bf p}_\Lambda
+ \beta_{\Xi}{\bf P}^*_{\Xi}\times\hat{\bf p}_\Lambda
+ \gamma_{\Xi} \hat{\bf p}_\Lambda\times\left({\bf P}^*_{\Xi}\times\hat{\bf p}_\Lambda\right)}{1+\alpha_{\Xi}{\bf P}^*_{\Xi}
 \cdot\hat{\bf p}_\Lambda} ,
\end{equation}
\end{widetext}
where $\hat{\bf p}_\Lambda$ is the unit vector of the $\Lambda$ momentum in the $\Xi$ frame.  
 
In the rest frame of the $\Xi$, the angular distribution of the $\Lambda$ is: 
\begin{equation}
\frac{\di N}{\di \Omega} = \frac{1}{4\pi} \left(1+\alpha_{\Xi^-} {\bf P}^*_\Xi \cdot \hat{\bf p}_\Lambda \right) ,
\end{equation}
As we have seen, rotational symmetry demands that the mean, momentum averaged ${\bf P}^*_\Lambda$ is
proportional to ${\bf P}^*_\Xi$ according to eq.~(\ref{linear}). Therefore we can obtain the relevant
coefficient $C$ by integrating (\ref{poladetail}) along the direction of ${\bf P}^*$ taken as
$z$ direction, weighted by the above angular distribution:
\be
 C_{\Lambda\,\Xi} = \int \di \Omega, \frac{\di N}{\di \Omega} {\bf P}^*_{\Lambda}\cdot \frac{{\bf P}^*_{\Xi}}{P^*_{\Xi}} 
 = \tfrac{1}{3}\left(2\gamma_{\Xi}+1\right) .
\ee
Using the measured~\cite{Agashe:2014kda} values for $\gamma_{\Xi^-}$ and $\gamma_{\Xi^0}$, the polarization transfers 
(which are the same as spin transfers, since  $S_{\Xi}=S_{\Lambda}$) are: 
\begin{align}
C_{\Lambda\Xi^-}&=\tfrac{1}{3}\left(2\times0.89+1\right)=+0.927 \nonumber \\ 
C_{\Lambda\Xi^0}&=\tfrac{1}{3}\left(2\times0.85+1\right)=+0.900
\end{align}
%



\end{document}